\documentclass[conference]{IEEEtran}

\usepackage{balance}

\IEEEoverridecommandlockouts 
\usepackage[pdftex]{graphicx}
\usepackage{algorithmic}
\usepackage{algorithm}
\usepackage{listings}
\usepackage[cmex10]{amsmath}
\usepackage{url}
\usepackage{multirow}

\title{Automatized Generation of Alphabets of Symbols}
\author{\IEEEauthorblockN{Serhii Hamotskyi\IEEEauthorrefmark{1}, Anis Rojbi\IEEEauthorrefmark{2}, Sergii Stirenko\IEEEauthorrefmark{1}, and Yuri Gordienko\IEEEauthorrefmark{1}}
\IEEEauthorblockA{\IEEEauthorrefmark{1}Igor Sikorsky Kyiv Polytechnic Institute, Kyiv, Ukraine, Email: shamotskyi@gmail.com}
\IEEEauthorblockA{\IEEEauthorrefmark{2}Laboratoire THIM (Technologies, Handicaps, Interfaces et Multimodalités), University Paris 8, Paris, France}
}

\begin{document}
\maketitle              

\begin{abstract}
In this paper, we discuss the generation of symbols (and alphabets) based on specific user requirements (medium, priorities, type of information that needs to be conveyed). A framework for the generation of alphabets is proposed, and its use for the generation of a shorthand writing system is explored. We discuss the possible use of machine learning and genetic algorithms to gather inputs for generation of such alphabets and for optimization of already generated ones.
The alphabets generated using such methods may be used in very different fields, from the creation of synthetic languages and constructed scripts to the creation of sensible commands for multimodal interaction through Human-Computer Interfaces, such as mouse gestures, touchpads, body gestures, eye-tracking cameras, and brain-computing Interfaces, especially in applications for elderly care and people with disabilities. 
\end{abstract}

\section{Introduction}
\IEEEoverridecommandlockouts\IEEEPARstart{T}{he need} to create writing systems has been with humankind since the dawn of time, and they always evolved based on the concrete challenges the writers faced. For example, the angular shapes of the runes are very convenient to be carved in wood or stone~\cite{williams1996origin}. The rapid increase of available mediums in the recent decades determined the need for many more alphabets, for very different use cases, such as controlling computers using touchpads, mouse gestures or eye tracking cameras.  It is especially important for elderly care applications~\cite{gordienko2017augmented} on the basis of the newly available information and communication technologies based on multimodal interaction through human-computer interfaces like wearable computing, augmented reality, brain-computing interfaces~\cite{stirenko2017user}, etc. 

Many approaches for the manual creation of alphabets have been used, but we are not familiar with a formalized system for their generation. Manually created alphabets are usually suboptimal. For example, it might be argued that the Latin alphabet favours the writer more than the
reader, since it evolved under the constraints of pen and paper, and those constraints are much less relevant in the computer age. Fonts which try to overcome this limitation exist~\cite{dotsies}. In a similar fashion, many systems do not use the possibilities given by the medium or context, electing to base themselves on already existing (familiar to the user, but suboptimal context-wise) symbols. A formalized framework capable of gathering requirements, generating symbols, grading them on a set of criteria and mapping them to meanings may be able to overcome many of those limitations.

The main aim of this paper is to propose a formalized framework capable of gathering requirements,
generating symbols, grading them on a set of criteria and mapping them to meanings, which potentially may overcome many of these limitations. \emph{The section II. Characteristics of a Rational Alphabet} contains the short characterization of basic terms and parameters of alphabets. The section \emph{III. Requirements for the needed alphabet} includes an example description of the requirements posed for alphabets used for shorthand systems. The section \emph{IV. Generation of
Glyphs} proposes
a method for the generation of glyphs with examples. The section \emph{V. Evaluation of Glyphs and Alphabets} contains discussion of fitness of glyphs/alphabets in relation to machine learning methods. The section \emph{VI. Discussion and future work} dedicated to discussion of the results obtained and lessons learned. 

\section{Characteristics of a rational alphabet}
"Glyph" is defined as unique mark/symbol in a given medium. "Symbol" is defined as a glyph with a meaning attached to it. "Alphabet" is defined as a system of such symbols, including possible modifiers and conventions.

Glyphs are generated and rated first, and meanings are assigned later; the alphabet as a whole is rated at the very end. This two-step process design choice is based on performance reasons (mutating individual glyphs and their meanings at the same time is too complex for any reasonably-sized alphabet) and is meant as a starting point for further research and adaptation. 



The following characteristics should generalize well for almost any alphabet, independently from the medium, dimensionality, and purpose. The vocabulary related to writing 2D characters with a pen or stylus is used, but this can be replaced with any other device.  

\subsection{Writing comfort and ergonomics}
For our purposes, we define comfort as "how easy and enjoyable is to use the alphabet". 
\begin{itemize}
        \item How much mental effort does the recall of the symbols require (ease of recall)
                \begin{itemize}
                        \item How familiar are the symbols to the user at the moment he is writing. 
                                \begin{itemize}
                                        \item Similarity to already known stimuli
                                        \item Availability of a mnemonic system
                                \end{itemize}
                \end{itemize}
        \item Fluency/flow, both for individual letters and their usual combinations.
        \item Physical limitations. For example, some strokes might be easier to write if someone is right-handed, or holds his pen in a certain way.
\end{itemize}

We suggest the following metrics as starting points for future research and discussion:

\subsubsection{Mental effort}
We think that this would be best measured via existing methods and some new methods of fatigue estimation on the basis of machine learning methods~\cite{gordienko2017ccp}. Changes in pupil size might be an especially interesting avenue in this aspect~\cite{Alns2014PupilSS}, as something objective and easy to measure.

If memory is more an issue than cognitive load, than generating the alphabet in such a way so that the glyphs can be "calculated" at writing time might help; as a very example of this, when we were manually creating our shorthand system, we decided to encode 
 time, modality, and person via a single glyph consisting of three parts. 
\subsubsection{Fluency}
Possible metrics for fluency could be:
\begin{itemize}
        \item Number of shap angles per glyph.
        \item Curvature per glyph. Both can be defined as sum the sum of absolute changes in direction per unit of distance.
        \item Ratio of strokes that mean something semantically, as opposed to "connecting one glyph with another", to the entire number.
        \item Number of easily connectable glyphs following each other in an average text, so that as little unnecessary movements are made. For example, given a representative source text, 
                \[c=\sum_{i=1}^n\sum_{j=1}^nE(g_i, g_j)P(g_i,g_j)\], where \(n\) is the number of existing glyphs, \(E(g_i, g_j)\) is how "easy" are the two glyph to connect, \(P(g_i, g_j)\) is how the probability \(g_i\) will be directly before \(g_j\).
\end{itemize}

\subsection{Writing speed}
Defined not as "how fast the pen moves", but rather "how much time is needed to convey the needed information".

\begin{itemize}
        \item How fast are individual glyphs to write. This intersects heavily with "Fluency".
                \begin{itemize}
                        \item Fluency from the subsection above.
                        \item How much the pen needs to travel to form the glyph.
                \end{itemize}
        \item How much "meaning" can be encoded in one glyph. This is directly related to redundancy and entropy, discussed in the following sections.
        \item The more simple glyphs should be mapped to the most common symbols.
\end{itemize}

A potentially interesting experiment would be timing people using the system, and dividing the amount of information written by the time taken; but this would raise questions about the input information. Accurately calculating the entropy of the conveyed information for this purpose would be practical only for alphabets used in very narrow and formalized contexts.

\subsection{Ease of recognition}

\begin{itemize} 
        \item How different are the glyphs between each other
        \item how much are distortions likely to worsen the recognition of the glyphs.
\end{itemize}

Additionally,  here various memory biases and characteristics of human memory will be at play (see, for example,the Von Restorff effect~\cite{hunt1995subtlety}).

\subsection{Universality}

Ideally, the glyphs should generalize well. That means that once learned for styluses,the same alphabet shouldn't be too hard to port to other mediums without losing many of the above mentioned characteristics. Excepting changes of dimensionality (3D-gestures might be hard to port to a 2D-stylus), this is probably the hardest to quantify and account for.

\section{Requirements for the needed alphabet}

Most writing systems have been heavily influenced by the constraints inherent in their area of use ---  purpose, characteristics of the information they needed to convey, materials. Even naturally evolving systems tend to converge towards local optima rather than a global optimum. Requirements and use patterns may gradually change, while the systems may be stuck in a state that is not optimal anymore. Therefore, a very careful analysis of the requirements and limitations is needed.

As example of applying our requirements above to our case of shorthand system, we can consider the following:
\begin{enumerate}
        \item On a purely symbolic level:
        \begin{enumerate}
                \item Writing letters
                \begin{enumerate}
                        \item number of strokes needed to encode individual letters
                        \item complexity of the resulting glyph
                \end{enumerate}
                \item Writing words
                \begin{enumerate}
                        \item connections between individual letters (glyphs)
                        \item how likely are letters that are easy to connect to each to be represented by easily connectable glyphs
                        \item if all existing glyphs are not identical in complexity, what is the ratio of easy-to-write glyphs to the complex ones in a typical text (the bigger the ratio, the better)
                \end{enumerate}
        \end{enumerate}
        \item Writing sentences:
        \begin{enumerate}
                \item are there any often-repeating words or groups of words which, when replaced by a shorter, even if complex, symbol, would lead to a gain in time? ("The" as a typical example).
        \end{enumerate}
        \item On a semantic level: Are there any grammatical categories or modalities that are represented in natural text with many letters, that when replaced by a single glyph or a modifier, would lead to a gain in time? (tenses, number, gender, hypotheticals, ...). The above mentioned symbol encoding time, modality, and person, to shorten words like "they would have been able to", happened at this level of abstraction.
        \item On an information theoretical level: How much redundancy is needed? How many errors in transcription can happen before the message becomes either unreadable or its meaning is distorted?  (Natural languages  are redundant via multiple mechanisms, notably via agreement in person, gender, case... Errors or interferences will still allow to understand what’s being said, up to a certain point. This may not be the case for constructed writing systems, if they
                are built with low redundancy.)~\cite{reza1961introduction} 
\end{enumerate}

One way to quantify some of the above would be analyzing source texts. At the end, at least the following information should be available:
\begin{itemize}
        \item frequencies of individual letters \(p_i\)
        \item most-needed connections \(c_{ij}\)
\end{itemize}

As example of how the information can be used, let's consider again our hypothetical shorthand system. Each of the generated glyphs can have three possible starting and ending strokes, represented by integers, and positioned at different heights.\(I_s, I_e=\{0, 1, 2\}\) Glyphs \(i, j\) where \(i_e=j_s\) are considered easily connectable. Using this information, later we can map the glyphs to meanings in such a way, that the letters that are most likely to follow each other are more likely to be
represented by easily connectable glyphs. The problem would be trivially solvable by having all glyphs start and end at the same point, but this would make it harder to differentiate the individual glyphs.

\section{Generation of the glyphs}
The second part of the proposed framework is the generation of possible glyphs. In this paper, Bezier curves have been used to generate the glyphs and calculate some of the needed metrics.
During the generation of the example glyphs, we made the following assumptions about the alphabet for which the glyphs are generated:
\begin{enumerate}
        \item The glyphs have a definite starting and ending point; the number of such points is limited, to facilitate connecting the symbols to each other.
        \item The stroke width does not vary (as, for example, in the case of Pitman shorthand), because of the low availability of pens able to convey even two levels of thickness and of low average penmanship skill in most people. (Though using it as a third or fourth dimension would certainly be possible.)
        \item The symbols will fit into a square bounding box.
\end{enumerate}
The generation of glyphs starts by fixing a definite starting and ending point and then adding a semi-random number of control points. Figures 1-3 are examples of glyphs generated using the above rules.

\begin{figure}[tbp]
\centering
        \includegraphics[width=0.75\hsize]{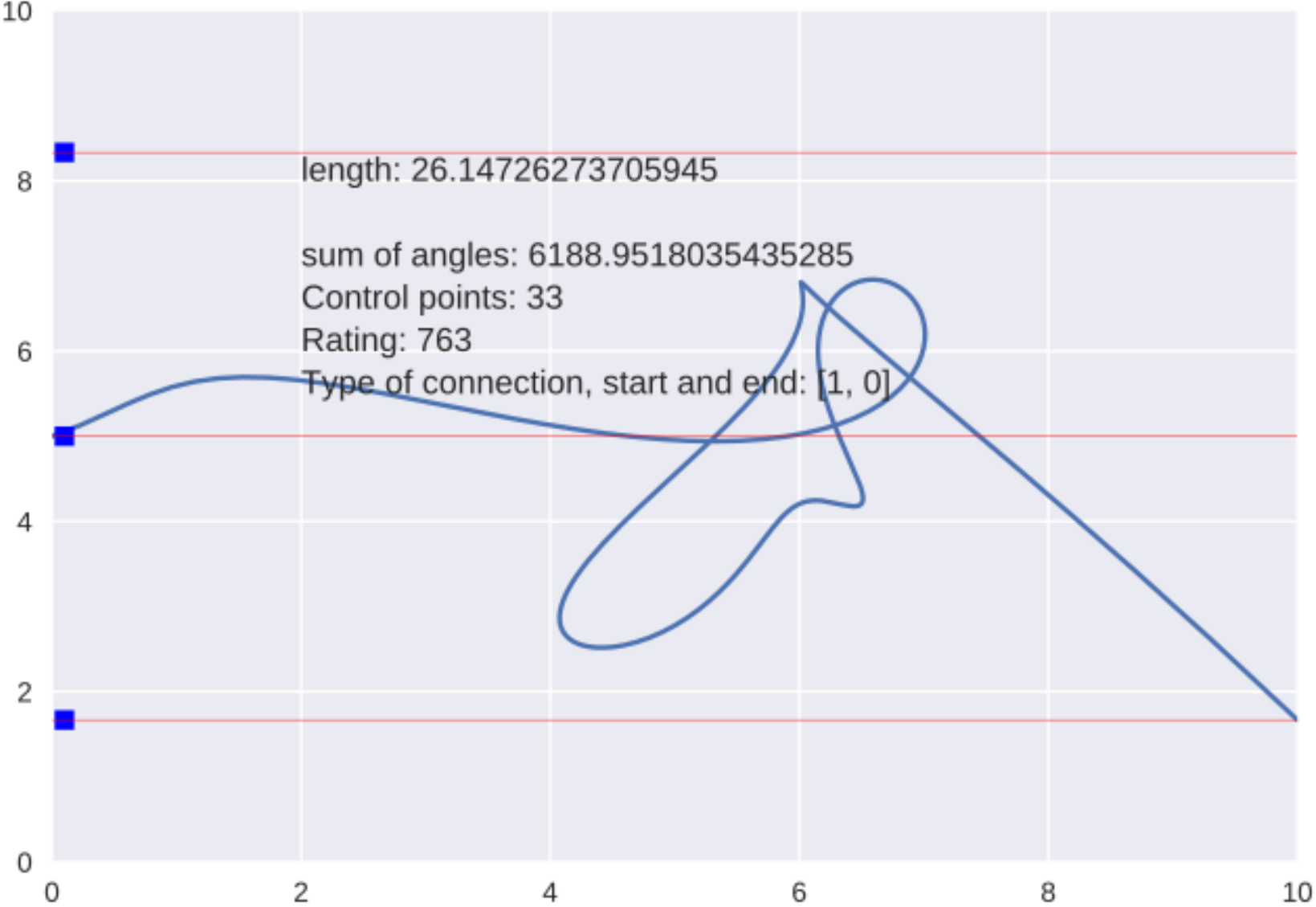}
\caption{Example of generated glyph with low fitness}
\end{figure}
\begin{figure}[tbp]
\centering
        \includegraphics[width=0.75\hsize]{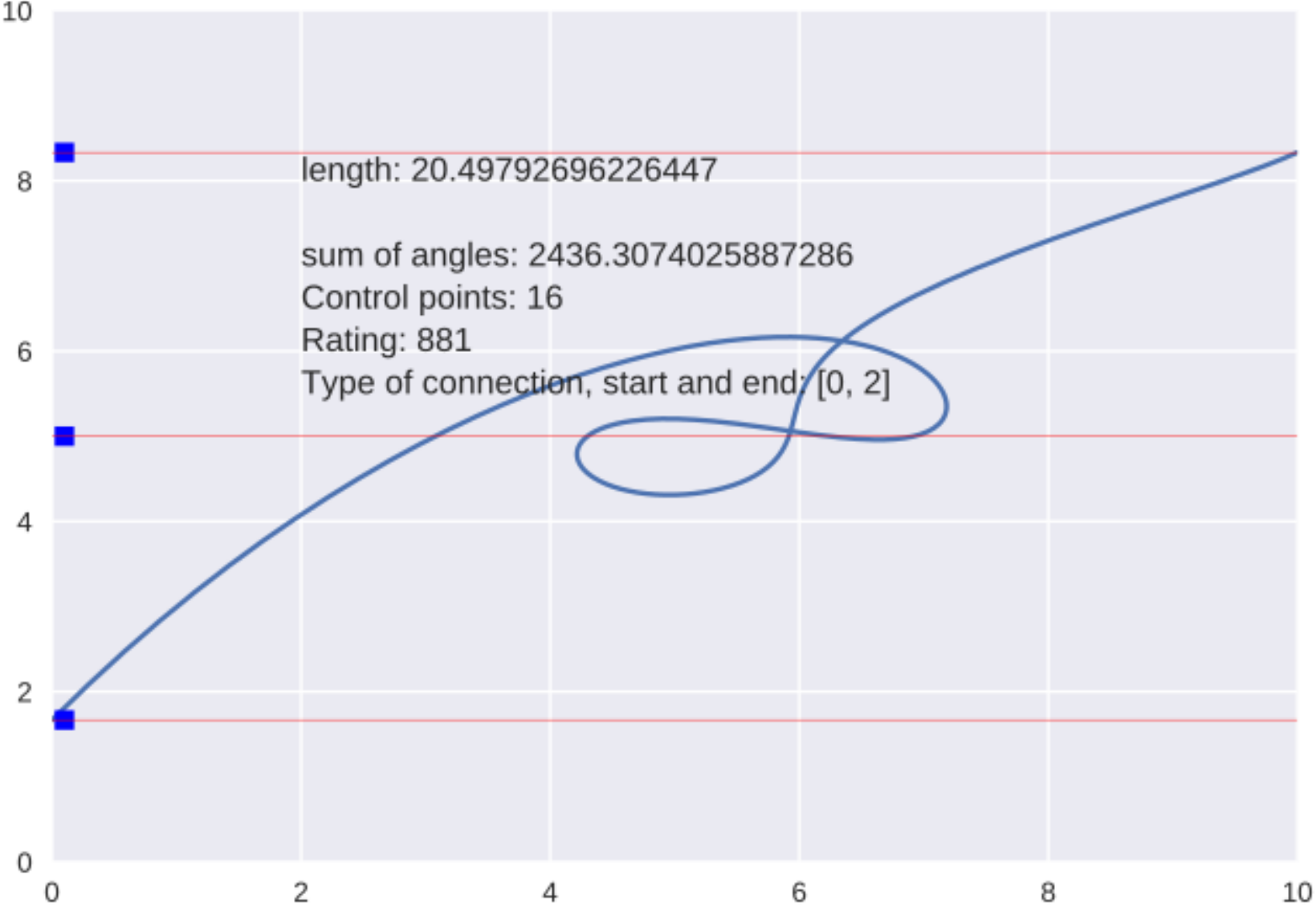}
\caption{Glyph with higher fitness      }
\end{figure}
\begin{figure}[tbp]
\centering
        \includegraphics[width=0.75\hsize]{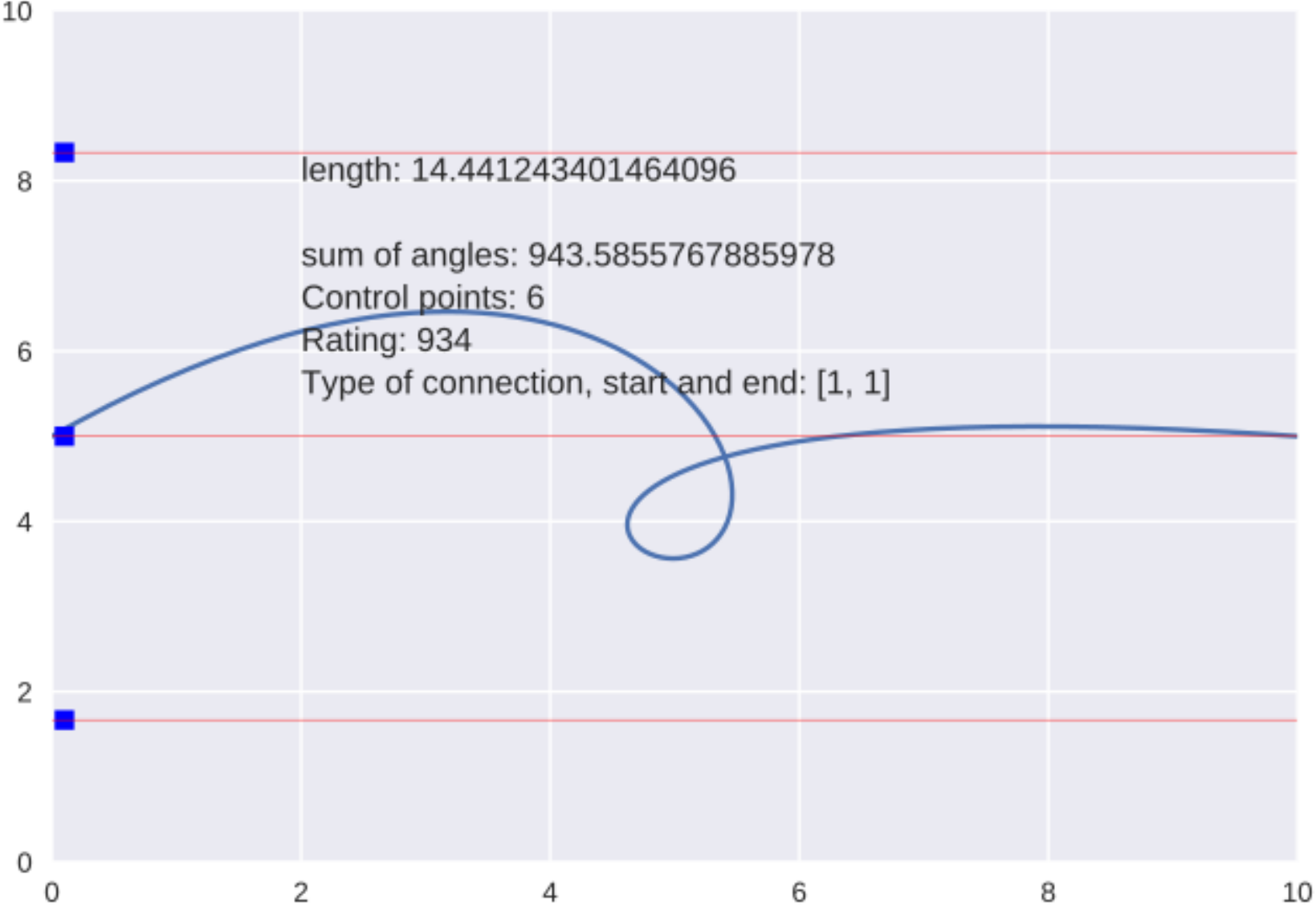}
\caption{The simpler a glyph is, the higher fitness it has}
\end{figure}

\balance

\section{Evaluation of Glyphs and Alphabets}
In this stage, the fitness of each glyph is determined. Many approaches are possible, and they heavily depend on the context and the medium for which the generation is being done. For our shorthand system, the main criteria were length and simplicity. The number of control points has been used as a proxy of fitness and has been partly accounted for in the generation phase (empirically, the more control points the more chaotic the glyph is). The second metric is complexity, which may be loosely
defined as "how hard it would be to write this symbol using a pen". For our purposes, complexity is defined as \(\frac{c}{l}\), where \(c\) is the sum of the angles in the polygonal representation of the curve (informally, how curved the glyph is; the more curves there are and the sharper the individual curves are, the bigger the value is), and \(l\) is the length of the curve (a certain amount of curves on a large glyph should not be penalized as much as the same amount on a smaller one). C is calculated by converting the curve between the first adjoining control points to a polygon, summing the absolute value of the angles between all adjoining lines, and repeating the process for all the successive control points.
\(c=\sum_{i=1}^n\sum_{j=2}^{p}L_n(j_i, j_i-1)\), where \(n\) is the number of control points,  \(p\) is the number of lines used to approximate the curve, L is the angle between two lines,  and \(j_i\) is the line after the control point \(i\). 

The reasons for defining \(c\) as we did are manifold, one of them being that a very similar metric is used for evaluating the similarity of the two glyphs to each other. Much better metrics are possible.

	The subjective reactions to signs might vary between people, differences due to age, cultural and/or language background are probable. This might be a promising area to study with the help of machine learning. Data like "Symbols similar to X perform poorly with demographic Y" would be valuable for creating alphabets when something about the probable users is known. 

Additionally, machine learning would open the doors for custom-tailored systems, where users rate some symbols and based on their feedback predictions are made about what other symbols they might like, remember and use. 
The first mapping of the generated glyphs, before its fitness is rated, is necessarily very tentative. In this paper we have not touched grammatical modalities and ways to shorten them in great detail, as they would merit quite a lot more research and space (and, probably, their own paper); regardless, they would have their place at this step of the framework. 
For an alphabet, our goals could be the following:
\begin{enumerate}
        \item As much high-fitness letters as possible
        \item Letters which are found the most often should have the highest fitness (that is, be as simple as possible).
        \item The letters should be unlike to each other
        \item The letters should be easily connectable
\end{enumerate}

The most important requirement is for the letters to be unlike each other. This is needed both for the resulting text to be readable (the existance of a 1-to-1 mapping between a text written in shorthand and a normal text, or at least for the resulting text being readable using contextual clues) and for improving the memorization of the glyphs (memorizing many similar stimuli is much harder than many different ones, unless a good framework for memorization is given, such as dividing symbols in parts). 

For our purposes histogram comparison was the most straight-forward to implement. The data for the histogram is provided by the angles computed at the previous step. Basic shapes and turns would be recognizable, and the difference between the two makeshift histograms would approximate the difference between the glyphs. Here, \(D_{ij}\) is the difference between glyphs \(i, j\).

Therefore, one formula for the fitness could be:
\[
f=\sum^{n}_{i=1}f_i+
\sum^{n}_{i=1}\sum^{n}_{i=1}D_{ij}+
\sum^{n}_{i=1}f_ip_i
\]
and the glyphs are picked so that the above formula is maximized. (The formula above does not include connections.)

A genetic algorithm at this point would attempt adding/removing/moving control points, switching glyphs between letters, introducing mirror-distortions etc. etc.

 \section{Discussion and future work}
 The basic ideas of this framework can be applied for the generation of any alphabet used in the real world. For touchpads, for example, connections may be built not using three possible endings, but 2D-points on the screen instead, and multitouch and weight-sensitivity may be included in the generation. By adding dimensions, 3D-gestures alphabets may be created. Much better heuristics for fitness may be created by more precise algorithms, machine learning and use of biology and cognitive science. The approaches demonstrated here are general enough to allow an enormous amount of flexibility in the kind of alphabets they may be used to create.
One of the more interesting avenues of further research would be creating algorithms for mapping glyphs to semantics, both to letters and to more complex grammar categories or structures. Finding (with AI?) the categories which could be shortened to one or two symbols is challenging by itself, but not all of the possible patterns found by an AI would be intuitive enough for a person to use or even to understand. 

\section*{Acknowledgment}
The work was partially supported by Ukraine-France Collaboration Project (Programme PHC DNIPRO) (http://www.campusfrance.org/fr/dnipro)

\bibliographystyle{myIEEEtran}


\end{document}